\documentclass[prl,twocolumn,amsmath,amssymb,10pt,aps,longbibliography,superscriptaddress,citeautoscript,bibnotes,footinbib]{revtex4-1}
\usepackage{draftcopy}
\usepackage{graphicx}
\usepackage{verbatim}
\usepackage{color} 
\usepackage{ulem}
\usepackage{amsmath}
\usepackage{algpseudocode}
\usepackage{mathrsfs}

\usepackage{amsmath}
\usepackage{bm}
\usepackage{latexsym}
\usepackage{graphics}
\usepackage{dsfont}
\usepackage{amssymb}
\usepackage{tabularx}
\usepackage{bbm}
\usepackage{bbold}
\usepackage{tikz}
\usetikzlibrary{quantikz}
\usepackage[colorlinks=true]{hyperref}   

\begin{document}
\title{Trotter error mitigation by error profiling with shallow quantum circuit}

\author{Sangjin Lee}
\thanks{Electronic Address: sangjin5190@gmail.com}
\affiliation{ Center for Quantum technology, Korea Institute of Science and Technology(KIST), Seoul, 02792, Republic of Korea}

\author{Youngseok Kim}
\thanks{Electronic Address: yskim87@gmail.com}
\affiliation{IBM Quantum, IBM Thomas J. Watson Research Center, Yorktown Heights, NY, USA}

\author{Seung-Woo Lee}
\thanks{Electronic Address: swleego@gmail.com}
\affiliation{ Center for Quantum technology, Korea Institute of Science and Technology(KIST), Seoul, 02792, Republic of Korea}

\date{\today}
 
\begin{abstract} 
Understanding the dynamics of quantum systems is crucial in many areas of physics, but simulating many-body systems presents significant challenges due to the large Hilbert space to navigate and the exponential growth of computational overhead. Quantum computers offer a promising platform to overcome these challenges, particularly for simulating the time evolution with Hamiltonians. Trotterization is a widely used approach among available algorithms in this regard, and well suited for near-term quantum devices. However, it introduces algorithmic Trotter errors due to the non-commutativity of Hamiltonian components. Several techniques such as multi-product formulas have been developed to mitigate Trotter errors, but often require deep quantum circuits, which can introduce additional physical errors. In this work, we propose a resource-efficient scheme to reduce the algorithmic Trotter error with relatively shallow circuit depth. We develop a profiling method by introducing an auxiliary parameter to estimate the error effects in expectation values, enabling significant error suppression with a fixed number of Trotter steps. Our approach offers an efficient way of quantum simulation on near-term quantum processors with shallow circuits.
\end{abstract}

\maketitle
Comprehending the behavior of quantum systems is essential for various branches of physics. In particular, describing the time evolution of many-body systems often requires navigating a large Hilbert space, posing significant challenges for conventional classical approaches; for example, the overhead grows exponentially with system size~\cite{ED1,ED2,ED3,ED4} or the accuracy can be compromised in approximate methods~\cite{DMRG,DMRG_white,DMRG2,DMRG3,MPS,nuemric1,MPS2,MPS3,TN1,TN2}. To address these challenges, quantum computer potentially provides a promising platform for quantum simulation as it inherently accommodates exponentially large Hilbert spaces~\cite{qsim_general,qsim_genengal2,qsim_general3, qsim_appl,qsim_appl2}.  

In order to simulate the dynamics governed by Hamiltonians $H$, it is desirable to approximate the evolution operator $e^{-iHt}$ accurately with quantum gates on a general-purpose quantum computer. Among various algorithms developed for that purpose~\cite{qsim_algo_LCU,qsim_algo_LCU2,qsim_algo_QSP,qsim_algo_quantumwalk,qsim_algo_quantumwalk2,qsim_algo_quantumwalk3,qsim_algo_qubitization,qsim_algo_qwalk,qsim_algo_random,qsim_algo_sym}, Trotterization, or product formula, is one of the most practically appealing methods for simulating many-body dynamics. This approach approximates $e^{-iHt}$ as a product form of quantum gates, each corresponding to the evolution of a divided part of the Hamiltonian~\cite{Suzuki1,Suzuki2,PF_ruth,PF_ruth2,PF_random,PF_random2,PF_second}. This approximation, however, introduces an {\it algorithmic} error due to the non-commutativity of each Hamiltonian component, which is referred to as the Trotter error. Increasing the number of divided steps reduces the Trotter error, but also increases the length of required quantum circuits, imposing a trade-off between the simulation accuracy and the circuit depth.

Numerous efforts have been made to enhance the accuracy of Hamiltonian simulation with limited-depth quantum circuits. The multi-product formula (MPF) is one of the most promising algorithms~\cite{MPF,MPF2,MPF3,MPF_extrapol}, which employs additional circuits to mitigate the Trotter error based on Richardson extrapolation~\cite{Richard}. While the original algorithm requires a challenging implementation including a linear combination of circuits~\cite{qsim_algo_LCU}, more practical methods have since been developed~\cite{MPF,MPF2,MPF3,MPF_extrapol}. In these methods, expectation values can be obtained with enhanced accuracy by running the same simulation multiple times with a different number of Trotter steps and classical post-processing~\cite{MPF}. However, increasing the number of quantum gates and modifying Trotter steps in deeper circuits would pose experimental challenges and introduce additional physical errors, still hindering practical quantum simulations on current and near-term quantum processors. 

In this letter, we propose a resource-efficient scheme to mitigate the Trotter error in Hamiltonian simulations. To enhance the accuracy of simulation with a limited number of quantum gates, we develop a profiling method to estimate the effect of Trotter errors in expectation values. Based on the profiled data, the Trotter error can be substantially suppressed without the additional cost of quantum gates. As being implementable with a fixed number of Trotter steps, our method does not introduce an additional effect of physical errors on the results. We believe that our profiling method provides an efficient tool to suppress the Trotter errors in Hamiltonian simulations with shallow quantum circuits in near-term quantum processors.

\textit{Trotter error profiling.}---We start by introducing a profiling method to mitigate the Trotter error in Hamiltonian simulation. For a given Hamiltonian $H$, the ideal evolution operator is written by $U(t)=e^{-i H t}$. We assume a Trotterized quantum circuit to approximate $U(t)$ up to $\left(\alpha-1\right)$-th order, which can be represented by
\begin{equation}
\label{eq:Voperator}
V(t) = U(t) + \sum_{s=\alpha}^\infty E_{s}t^s, 
\end{equation}
with the Trotter error operators $\{ E_s\} $. 
For example, for a given Hamiltonian $H=H_1+H_2$, the operator $V(t)=e^{-i H_1 t}e^{-i H_2 t}$ approximates the ideal evolution operator $U(t)=e^{-i (H_1+ H_2)t }$ up to the first order. The gap between the ideal $U(t)$ and approximated operator $V(t)$ thus introduces the Trotter error, which can then be represented by the operators $E_2= -\frac{1}{2} [H_1,H_2]$, $E_3=\frac{-i}{3!}\left(H_1[H_2,H_1]+[H_2,H_1^2]+[H_2,H_1]H_2+[H_2^2,H_1] \right)$, and so on. 

We aim to analyze and mitigate the effect of errors corresponding to $\{E_s\}$ in the expectation value of the Hamiltonian simulation. For a given Trotterized circuit implementing $V(t)$, we can always consider its inverse quantum circuit implementing $V^\dagger(t)$ to satisfy $V^\dagger(t) V(t)=\mathbb{1}$. The circuit $V^\dagger(t)$ can be constructed by reversing the sequence of quantum gates and applying phase rotations with opposite signs. For example,
\begin{tikzpicture}
\node[scale=0.67] {
$V(t)=$
\begin{quantikz}
  &  \gate[2]{e^{-i t Z_{1} Z_2} } &\gate{R_x(2h t)}&\qw  \\
  &                                &\gate{R_x(2h t)}&\qw 
\end{quantikz}$\Rightarrow V^\dagger(t)=$
\begin{quantikz}
    &\gate{R_x(-2h t)}& \gate[2]{e^{+i t Z_{1} Z_2} } &\qw \\
    &\gate{R_x(-2h t)}&                               &\qw 
\end{quantikz}
};
\end{tikzpicture}
where $R_x(\theta)$ is a single-qubit rotation by $\theta$ along the $x$-axis. Implementing $V^\dagger(t)$ thus requires only the rearrangement of the gates used in $V(t)$ without additional resource cost regarding both circuit depth and complexity.  

The unitarity condition of $U(t)$ and $V(t)$ leads to the relation between $\{E_s\}$ and $\{E^\dagger_s\}$, which plays a crucial role in our method. From Eq.~(\ref{eq:Voperator}) and $U^\dagger(t)U(t) =\mathbb{1}$ and $V^\dagger(t)V(t) =\mathbb{1}$, we can obtain
\begin{equation}
\nonumber
U^\dagger(t)\sum_{s=\alpha}^\infty E_{s}t^s+\sum_{s=\alpha}^\infty E^\dagger_{s}t^s U(t)+\sum_{s,s'=\alpha}^\infty E^\dagger_{s}E_{s'}t^{s+s'}=0,
\end{equation}
the expansion of which for $t$ gives the condition between $\{ E_{s}\}$ and $\{ E^\dagger_{s}\}$, e.g., $E^\dagger_\alpha + E_\alpha=0$ at $\mathcal{O}(t^\alpha)$. For more details, see the supplemental material~\cite{SM}.

Building on these observations, we define a set of composite operators of $V(t)$ and $V^\dagger(t)$ as follows, 
\begin{equation} \label{eq:profile}
\begin{split}
V_{2N}^{(1)}(a,t) &= V_N\left(a t\right)V_N\left(\bar{a} t\right), \\ V_{2N}^{(2)}\left(a,t\right) &= V_N^\dagger\left(-a t\right)V_N\left(\bar{a} t\right),\\
V_{2N}^{(3)}(a,t) &= V_N\left(a t\right)V_N^\dagger\left(-\bar{a} t\right), \\ V_{2N}^{(4)}\left(a,t\right) &= V_N^\dagger\left(-a t\right)V_N^\dagger\left(-\bar{a} t\right), \\
\end{split}
\end{equation}
with an arbitrary parameter $a\in \mathbb{R}$ and $a+\bar{a}=1$. Here, $N$ indicates the number of iteration steps of the Trotterized circuit $V\left( \frac{t}{N} \right)$ over $t$, {\it i.e.}, $V_N(t)\equiv \left(V\left(\frac{t}{N}\right)\right)^N$.
Hereafter, we will formulate our method by setting $N=1$ for simplicity. We note that the generalization for arbitrary $N>1$ is straightforward by treating $V_N(t)$ as another Trotter circuit, say $\bar{V}(t)$.

In an {\it ideal} quantum simulation, $U(at)U(\bar{a}t)=U(t)$ as $a+\bar{a}=1$, ensuring that varying the parameter $a$ should not change the simulated results. On the other hand, when simulating on the Trotterized quantum circuits $V^{(i=1,2,3,4)}_{2}(a,t)$, the results are dependent on the parameter $a$. For example, the simulated data generated by $V_2^{(i=1,2,3,4)}(a,t)$ in Eq.~(\ref{eq:profile}) vary with different values of $a$, as the combination of the terms $E_s(at)^s$, $E_s(\bar{a}t)^s$, and their conjugations produce distinct Trotter error profiles. We leverage this property by using $a$ as a probe to profile the Trotter error landscape of $V_2^{(i)}(a,t)$. We will omit $i=1,2,3,4$ in superscripts for brevity hereafter. 

\begin{figure}[t!]
\includegraphics[width=.5\textwidth]{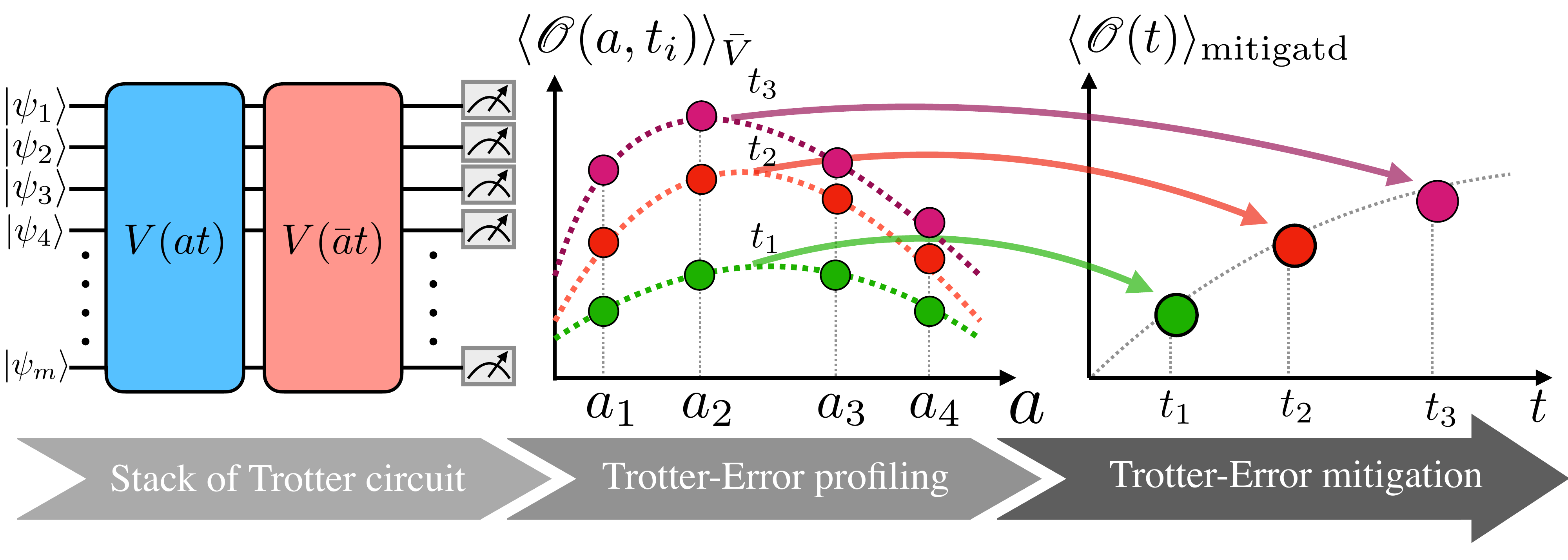}
\caption{Profiling method to mitigate Trotter error. (left) First, perform simulations on the Trotterized circuit implementing the composite operators in Eq.~(\ref{eq:profile}). (middle) To profile the effect of Trotter error, measure the expectation values of an observable by varying $a$ with fixed $t$. (right) Estimate the ideal expectation value by comparing the obtained data to theoretically expected behaviors.}
\label{fig1}
\end{figure}

Suppose that we perform a simulation to compute the expectation value of an observable $\mathscr{O}$ on a circuit $C(t)$ with an initial state $|\psi\rangle$. The expectation value of $\mathscr{O}$ obtained by simulation on this circuit can be written by $\langle \mathscr{O}  \rangle_{C(t)}= \langle   \psi |C^\dagger(t)  \mathscr{O}  C(t)|\psi \rangle$, while the expectation value by the ideal Hamiltonian evolution $U(t)$ is given by  $\langle \mathscr{O}\rangle_{U(t)}$. We define a measure of the Trotter error as
\begin{align*}
\varepsilon_{C(t)} = \langle \mathscr{O}  \rangle_{C(t)} -\langle\mathscr{O}  \rangle_{U(t)},
\end{align*}
for a Trotterzied circuit $C(t)$. The Trotterized quantum circuit $V_2^{(i)}(a,t)$ in Eq.~(\ref{eq:profile}) thus incurs Trotter error $\varepsilon_{V_2^{(i)}(a,t)}$, which can be represented as a function of the auxiliary parameter $a$ and $t$.  
For example, the Trotter error in the simulation on the circuit $V_{2}^{(1)}(a,t)= V\left(a t\right)V\left(\bar{a} t\right)$ is written by (see \cite{SM} for more details)
 \begin{align}
\varepsilon_{V_2^{(1)}(a,t)}=t^\alpha \langle  E^\dagger_\alpha \mathscr{O} +h.c. \rangle  \left(a^\alpha+ {\bar{a}}^\alpha\right) +  \mathcal{O}(t^{\alpha+1}). \nonumber
 \end{align}
{\it A priori}, computing the matrix elements that appear in $\varepsilon_{V_2^{(i)}(a,t)}$, such as $\langle  E^\dagger_\alpha \mathscr{O} +h.c. \rangle$, is challenging for general many-body system. 

Let us now show that how these matrix elements can be estimated by profiling the landscape of Trotter errors through measurements of $\langle \mathscr{O} \rangle_{V_2^{(i)}(a,t)}$ for various values $a$ at a given time $t$. 
In order to demonstrate this, we define the averaged simulated data over $i=1,2,3,4$,
\begin{align}
\label{theoretic}
\langle \mathscr{O} \rangle_{\bar{V}(a,t)} \equiv \frac{1}{4}\sum_{i=1}^4\langle \mathscr{O}\rangle_{V_2^{(i)}(a,t)}=\langle \mathcal{O}\rangle_{U(t)} + \bar{\varepsilon}(a,t),
\end{align}
where $\bar{\varepsilon}(a,t)=\sum_{s=\alpha} \bar{\varepsilon}_{s}(a) t^s$ and $\bar{\varepsilon}_{s}(a)$ is the averaged Trotter error of order $s$ in $t$.
For example, the $\alpha$-th order error is $\bar{\varepsilon}_{\alpha}(a)=m_{\alpha}  (a^\alpha + {\bar{a}}^\alpha)$, where 
$m_\alpha = \langle \left(E^\dagger_{\alpha} +(-1)^\alpha E_\alpha\right) \mathscr{O} +h.c.  \rangle $ is a matrix element determined by the given Trotterized quantum circuit $V(t)$. Likewise, higher-order Trotter errors can be represented with corresponding higher-order matrix elements~\cite{SM}.

By collecting a set of simulated data
\begin{align}
\{ \left(a_{1}, \langle \mathscr{O}\rangle_{\bar{V}(a_1,t)} \right), \left(a_{2}, \langle \mathscr{O}\rangle_{\bar{V}(a_2,t)}  \right) \cdots \},
\label{data}
\end{align} 
and comparing them to the theoretical expectation from Eq.~\eqref{theoretic}, we can estimate the matrix elements $m_\alpha$. 
If we figure out the value of $m_\alpha$, the Trotter error $\bar{\varepsilon}(a,t) = \langle \mathscr{O}  \rangle_{\bar{V}(t)} -\langle\mathscr{O}  \rangle_{U(t)}$ can be mitigated up to $\mathcal{O}(t^{\alpha})$. 
Note that this procedure can be understood as a least square fitting process, where the profiled data in Eq.~(\ref{data}) is fitted into the calculated function in Eq.~(\ref{theoretic}). Figure~\ref{fig1} illustrates the summary of our proposal.

We remark on the circuit cost of the proposed Trotter error profiling method.
To extract the ideal expectation value $\langle \mathscr{O}\rangle_{U(t)}$ up to $\mathcal{O}(t^{\alpha+n})$, the protocol requires sampling $2n+1$ different values of $a$ to determine the matrix elements in the Trotter error $\bar{\varepsilon}(a,t)$. This follows from the fact that the hierarchical relations between $\{ E_s\}$ and $\{ E^\dagger_s\}$ imposed by the unitarity condition restricts the number of independent variables (see details in \cite{SM}). 
To estimate the matrix elements, the number of independent variables should be equal to the number of independent equations constructed by Eq.~\eqref{theoretic} and \eqref{data}.
In this manner, we can find that the proposed profiling method is able to mitigate the Trotter error up to $\mathcal{O}\left(t^{2\alpha-2}\right)$, details of which is explained in the supplemental material~\cite{SM}.
This implies that improving performance of our method requires the use of a higher-order Trotter formula.

Let us now compare our profiling method with the MPF~\cite{MPF,MPF_extrapol,MPF2,MPF3} (see also the brief review of MPF in \cite{SM}).
MPF method allows us to mitigate the Trotter error up to $\mathcal{O}(t^{\alpha+N-1})$ by extrapolation with the data obtained from Trotterized circuits $V_{s}(t)= \left(V\left(\frac{t}{s}\right)\right)^s $, $(s=1,\cdots, N)$. By using the symmetic Trotter formula such that $V(-t)=V^\dagger(t)$, this can be further improved up to  $\mathcal{O}(t^{\alpha+2N-1})$~\cite{MPF_extrapol,SM}.
Thus, there exists a critical number of iteration steps for the Trotterized circuit, $N_c$, which divides beneficial regions between our profiling method and MPF. This critical number is determined by the mitigation limit of both methods, 
\begin{align}
\text{(regular)}&~~
2\alpha-2 = \alpha+N_c-1 \Rightarrow 	N_c=\alpha-1,\nonumber\\
\text{(symmetric)}&~~
2\alpha-2 = \alpha+2N_c-1 \Rightarrow 	N_c=\frac{1}{2}\left(\alpha-1\right). ~\nonumber
\end{align}
Therefore, the Trotter error mitigation strategy should be appropriately chosen based on the feasible value of $N$.

\textit{Applications.}--
We now apply our profiling method for simulating two typical models, i.e., the one-dimensional transverse field Ising model (1$D$ TFIM) and $XXZ$ chain.

(i) 1$D$ TFIM: Let us consider a simple 1$D$ TFIM with the Hamiltonian given by, 
\begin{align}
H_{\text{TFIM}} = J\sum_{i=1}^3 Z_i Z_{i+1}  + h \sum_{i=1}^4 X_i, \nonumber
\end{align}
where $J$ represents the spin-spin strength, and $h$ denotes the external transvers field. The ideal time-evolution operator is given by $U(t)=e^{-i H_{\text{TFIM}} t}$. To illustrate our scheme within this model, we fix the parameters as $J=1$ and $h=1/3$.

In order to simulate the model through quantum circuit, suppose that we trotterize $U(t)$ into 
\begin{align*}
V_3(t) &= \prod_{i=1}^{3} V_{ZZ}(p_i t) V_X(q_i t),\\
V_{ZZ}(t) &=\prod_{i\in \text{odd}} \left( e^{-i Z_{i}Z_{i+1} t}\right) \prod_{i\in \text{even}} \left( e^{-i Z_{i}Z_{i+1} t}\right),\\
V_X(t)&= \prod_{i=1}^4 \left( e^{-i X_i t/3}\right) ,
\end{align*}
with $(p_1,q_1,p_2,q_2,p_3,q_3)=\left(\frac{7}{24},\frac{2}{3},\frac{3}{4},-\frac{2}{3},-\frac{1}{24},1\right)$, which is the third-order Trotter-formula known as the Ruth formula \cite{PF_ruth,PF_ruth2}.
We can also consider the higher order formula, e.g., the forth order ($V_{\alpha=5}(t)$) Suzuki-Trotter formula \cite{Suzuki1,Suzuki2}. Note that $(V_{\alpha=5}(-t))^\dagger=V_{\alpha=5}(t)$, {i.e.} symmetric formula so that Eq.~\eqref{eq:profile} reduces to a single composite operator $V_{\alpha=5}(a t)V_{\alpha=5}(\bar{a}t)$.

Assume that the target observable to simulate is written by
\begin{align}
\mathscr{O}= \frac{1}{4} \sum_{i=1}^4 X_i +\frac{1}{3} \sum_{i=1}^3 Z_iZ_{i+1}, \nonumber
\end{align}
with respect to an initial state
\begin{align}
|\psi \rangle = \frac{1}{2}
\begin{pmatrix}
1\\
0
\end{pmatrix}
\otimes
\begin{pmatrix}
1\\
i
\end{pmatrix}
\otimes
\begin{pmatrix}
1\\
1
\end{pmatrix}
\otimes
\begin{pmatrix}
0\\
1
\end{pmatrix}.\nonumber 
\end{align}
We then simulate $\langle \mathscr{O}\rangle_{\bar{V}(a,t)}$ defined in Eq.~(\ref{theoretic}) with each Trotter-formula, and mitigate the Trotter errors by the profiling method introduced in the previous section. 
We can calculate the mitigated Trotter error as 
\begin{align}
\varepsilon_{\text{error}}(t) = \Big| \langle \mathscr{O} \rangle_{\tilde{U}(t)} - \langle \mathscr{O}\rangle_{U(t)}   \Big|, 
\end{align}
where $\langle \mathscr{O}\rangle_{\tilde{U}(t)}$ represents the expectation value obatained by the proposed profiling method and $\langle \mathscr{O}\rangle_{U(t)}$ is the ideal expectation value computed by the numerical diagonalization.

\begin{figure}[t!]
\includegraphics[width=.5\textwidth]{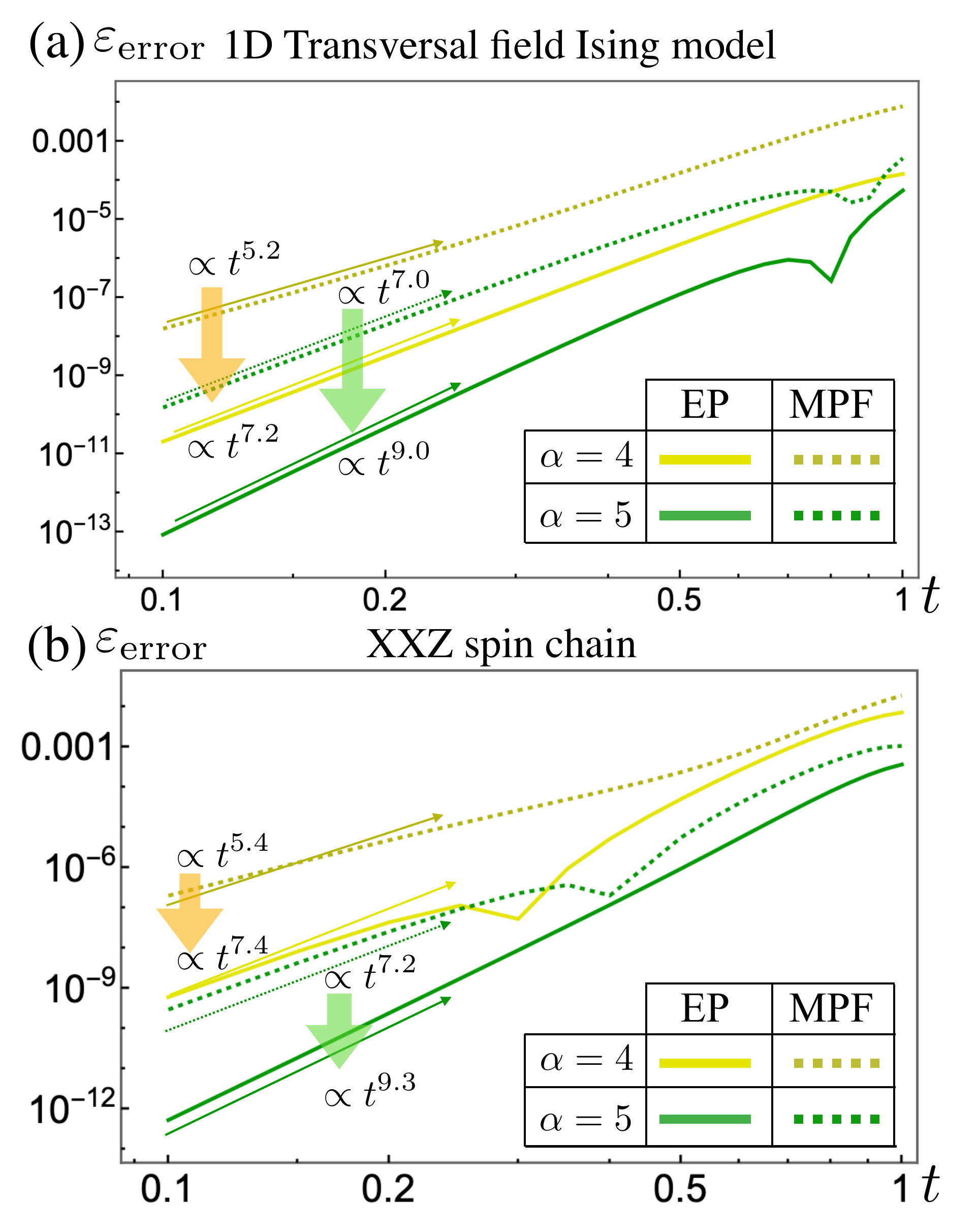}
\caption{Trotter errors $\varepsilon_{\text{error}}(t)$ mitigated by the profiling (EP, solid lines) and MPF(dotted lines) methods as a function of time $t$ with different Trotter-formulae, $\alpha=4$(yellow), $5$(green) for (a) the 1D transverse field Ising model (1$D$ TFIM) and (b) the $XXZ$ model. The leading temporal behaviors of $\varepsilon_{\text{error}}(t)$ are estimated by the gradient fitting in the $\log$-$\log$ plot. We overlaid the guidelines of gradients for eyes (thin lines). For MPF method, we used two units of Trotter-circuits for a comparison to Trotter-error profiling method. 
The results obtained through our Trotter-error profiling method demonstrate superior performance compared to the MPF method, exhibiting improvements in both the order-of-magnitude accuracy and the temporal scaling behavior of the error.}
\label{fig2}
\end{figure}

Figure~\ref{fig2}(a) shows the Trotter errors mitigated by the profiling method (EP) and the MPF method for different values of $\alpha$ as the evolution time $t$ varies. The results demonstrate that our profiling method outperforms MPF for a given $\left(\alpha-1\right)$-th order Trotterized-circuits. To quantitatively analyze the temporal dependence of $\varepsilon_{\text{error}}(t)$, we perform gradient fitting in a $\log$-$\log$ plot for both methods. The extracted values align well with our expected theoretical behavior.
In the supplemental material~\cite{SM}, we also compare $\varepsilon_{\text{error}}(t)$ with both the profiling (EP) and MPF methods as increasing $N$ with fixed $t$ for various Trotter-formulae~\cite{SM}.

(ii) $XXZ$ spin chain :  
As an another example, we consider $XXZ$ chain related to 1D Fermi-Hubbard model by the Jordan-Wigner transformation \cite{JWtransf,FHdual}, 
\begin{align*}
H_{XXZ} = J\sum_{i=1}^3  \left(X_{i}X_{i+1} +Y_i Y_{i+1} +\eta Z_{i}Z_{i+1}\right),
\end{align*}
where $\eta$ controls anisotropy of the model. We here set $J=1$ and $\eta=1/3$ for an illustration.
To simulate the ideal time-evolution operator $U(t) =e^{- i H_{XXZ} t}$, suppose that we trotterize $U(t)$ as
\begin{align*}
V_3(t) &= \prod_{i=1}^3 \left( V_{1,2}(p_i t)V_{3,4}(p_it) V_{2,3}(q_it)\right),\\
V_{i,i+1}(t)&= e^{- i  \left(X_{i}X_{i+1} +Y_i Y_{i+1} +\frac{1}{3}Z_{i}Z_{i+1}\right)t},
\end{align*}
where $V_3(t)$ is given as the same Ruth formula~\cite{PF_ruth,PF_ruth2} used in the previous example ($1D$ TFIM). We can also generate the fourth-order Trotter-formulae following Suzuki-Trotter formula.

We consider the time-evolution of an observable
\begin{align}
\mathscr{O} =\frac{1}{4} \sum_{i=1}^4 Z_i, \nonumber
\end{align}
with respect to an initial state
\begin{align}
|\psi \rangle = \frac{1}{2}
\begin{pmatrix}
1\\
0
\end{pmatrix}
\otimes
\begin{pmatrix}
1\\
i
\end{pmatrix}
\otimes
\begin{pmatrix}
1\\
1
\end{pmatrix}
\otimes
\begin{pmatrix}
0\\
1
\end{pmatrix}.\nonumber 
\end{align}

By simulating $\langle \mathscr{O}\rangle_{\bar{V}(a,t)}$, we can mitigate the Trotter errors by the proposed profiling method.
In Fig.~\ref{fig2}(b), we plot the mitigated Trotter errors $\varepsilon_{\text{error}}(t)$ mitigated using both the profiling method (EP) and MPF method with two given Trotter-formulae. Likewise in the previous example, we evaluate the performance of Trotter error mitigation by analyzing the power-law dependence of $\varepsilon_{\text{error}}(t)$ for time $t$, extracted from the $\log$-$\log$ plot. Figure \ref{fig2}(b) shows that our profiling method (EP) can achieve superior Trotter error mitigation than MPF. We also compare the $N$ dependence of both methods in the supplemental material~\cite{SM}.

\textit{Discussions.}--
We have proposed a resource-efficient method for mitigating Trotter errors in Hamiltonian simulations. By profiling the impact of Trotter errors in expectation values, our approach significantly suppresses these errors without incurring additional quantum gate costs. This method employs two stacks of Trotter circuits with an auxiliary parameter $a$, whose variation does not affect the ideal quantum simulation but influences the Trotterized quantum circuit. This enables us to map the landscape of Trotter errors effectively. Applying our profiling method to the simulation of two representative models, i.e.,~the one-dimensional transverse field Ising (1$D$ TFIM) and $XXZ$ spin chain, we have demonstrated that it can suppress Trotter errors by approximately two orders of magnitude more than MPF for $\alpha=4,5$ order formulae.

For a given physical noise in $N$ stacks of Trotter circuits in noisy hardware, MPF method requires $\mathcal{O}(N^2)$ number of Trotter circuits for error mitigation, while our profiling method requires only $\mathcal{O}(N)$~\cite{SM}. Moreover, our profiling method employs shallower quantum circuits for a given mitigation order. This circuit shallowness enhances the effectiveness of known error mitigation techniques, such as zero-noise extrapolation~\cite{ZNE1,ZNE2} and probabilistic error cancellation~\cite{PEC1,PEC2}, potentially improving the overall quality of quantum simulation results. Therefore, our profiling method offers a promising approach to addressing both algorithmic and physical errors in quantum simulations.

We have observed that exploiting the hierarchical relationships between matrix elements in the profiled errors can enhance the performance of Trotter error mitigation by reducing the number of independent variables that need to be determined~\cite{SM}. This suggests that further exploring these hierarchical relationships using physical properties of $H$, such as the symmetry of a given Hamiltonian,  could provide additional opportunities to improve the performance of our error-profiling method.

This research was funded by Korea Institute of Science and Technology (2E33541) and National Research Foundation of Korea (2022M3K4A1094774). 

\bibliography{TEMref}

\end{document}